\documentclass[aps,pre,amsmath,twocolumn,showpacs]{revtex4}
\usepackage{amsmath}
\usepackage{amsfonts}
\usepackage{graphicx}
\usepackage{hyperref}
\usepackage{mathtools}
\usepackage{verbatim}
\usepackage{epstopdf}
\usepackage{subfigure}
\usepackage{hyperref}
\usepackage{latexsym}
\usepackage{dcolumn}
\usepackage{epsf}
\usepackage{float}
\DeclarePairedDelimiter\abs{\lvert}{\rvert}

\usepackage{color}

\begin{document}
\title{Explosive synchronization in phase-frustrated multiplex networks}
\author{Pitambar Khanra$^1$}
\author{Prosenjit Kundu$^1$}
\author{Chittaranjan Hens$^2$} 
\author{Pinaki Pal$^1$}
\affiliation{$^1$Department of Mathematics, National Institute of Technology, Durgapur 713209, India,\\ $^2$Physics \& Applied Mathematics Unit, Indian Statistical Institute,  Kolkata 700108, India}

\begin{abstract}
We investigate the phenomenon of first order transition (explosive synchronization (ES)) in adaptively coupled phase-frustrated bi-layer multiplex network. We consider Sakaguchi-Kuramoto (SK) dynamics over the top of multiplex networks 
and we establish that ES can emerge in all layers of a multiplex network even when one of the layers may not exhibit ES in  the absence of the inter-layer connections. We clearly identify the regions of the parameter space, in which the multiplexity wins over frustration parameter and network structure for the emergence of ES. Based on the mean field analysis around the coherent state  and a perturbative approach around the incoherent state we analytically derive the synchronization transition points (backward and forward) of all layers of multiplex network as well as its mono-layer counterpart satisfying a close agreement  with the numerical  results.
\end{abstract}

 \maketitle
It has been widely accepted that transition to synchronization in networks of coupled oscillators is a continuous process \cite{Boccaletti2002,Pikovsky2003,Strogatz2003,Kuramoto1984,Bonilla2005} until the discovery of irreversible or discontinuous  synchronization transition~\cite{Pazo2005}.  Subsequently, discontinuous or first order transition to synchronization (also known as explosive synchronization (ES)) has  been extensively explored in networks of coupled oscillators~\cite{Jesus2011,Kurths2016,Leyva2013,Boccaletti2016}. The practicality of ES has been tested in acoustical signal transduction in the cochlea  (modeling the hair cell) \cite{Wang2017}, hypersensitivity in FM brains \cite{Lee2018} and experimentally verified in mercury beating-heart (MBH) oscillators \cite{Kumar2015}. Researchers reported that ES  can be achieved  by correlating the netural frequencies with the degrees of the networks \cite{Jesus2011,Pinto2015} or by designing a frequency based  weighted coupling \cite{Zhang2013,Leyvaweighted2013, Xu2016}. 
 
 However, these pre-conditions for the emergence of ES in a network of oscillators can be waived by setting an adaptive factor in the coupling term where adaptation is extracted from the global order parameter of the  phase oscillators \cite{Filatrella2007}. Recently, adaptive strategy has also been extended in complex networks (including multiplex networks) of phase oscillators \cite{Zhang2015,Danziger2016}. The growing interest among the researchers to investigate dynamical processes in multiplex networks~\cite{Kachhvah2017,Nicosia2017}  may be attributed to the fact that diverse complex systems ranging from engineering to transportation or to ecology can be mapped to multiplex networks~\cite{Domenico2016}. Moreover, due to the presence of layer-layer interaction in multiplex networks, some nontrivial effect on different dynamical phenomena is expected. For instance, the role of layer-layer interaction in multiplex networks on the dynamical process of diffusion~\cite{Gomez2013}  can be faster in multiplex network than its monolayer counterpart due to layer-layer interaction. 

Here we have focused on adaptively coupled multiplex network, expecting that the presence of  layer-layer interaction may induce  unexpected effect on the synchronization transition. We have considered either $(1)$ each layer is  structurally different than other i.e heterogeneity of the degree distributions are not identical although both layers consist same dynamics (identical frustration and frequencies are drawn from the same distribution) or  $(2)$ both the networks are structurally equivalent but dynamically different  (frustration terms are not identical). Note that, for both cases, one of the layer shows continuous transition in the absence of inter layer interaction.
\par The key question, we ask here, whether ES can be established in all layers of a multiplex network in which one of the layers may not exhibit ES in absence of inter layer connection?  We have tried to answer the question by exploring a multiplexed (bi-layer) phase-frustrated system \cite{Sakaguchi1986} called as Sakaguchi-Kuramoto (SK) dynamics. We have shown that an adaptive multiplex network can indeed exhibit ES in all phase-frustrated layers. More precisely our paper establishes that fact (with proper analytical treatment) that depending on the network structure  and frustration parameter, a monolayer may not exhibit ES  but it does guarantee ES in multilayer structure if at least one of its layer is in the regime of ES, an interesting phenomena not explored    earlier (in frustrated environment) to the best of our knowledge. For analytical treatment, we have used an annealed network based approximation \cite{Ichinomiya2004,Coutinho2013,Kundu2017} and perturbative approach around the  incoherent state \cite{Wu2018} to find the backward and forward transition point of the hysteresis loop. 

\begin{figure}[h]
\includegraphics[height=!,width=0.48\textwidth]{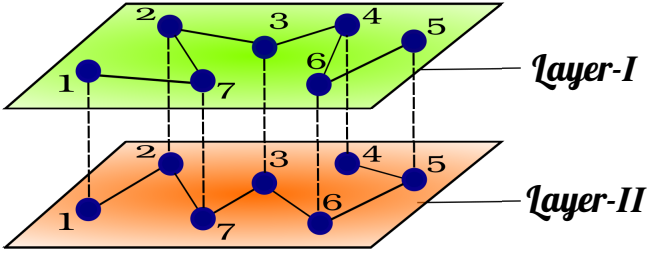}
\caption{Schematic diagram of a multiplex network consisting of two layers of different network topologies.}
\label{schematic_diagram_multilayer}
\end{figure}

We consider two complex networks of same size $N$ which form a multiplex network if there is intra-layer connections. A schematic diagram of a multiplex network in presence of a special type of intra-layer connections is shown in Fig.~\ref{schematic_diagram_multilayer}. Thick black lines (shown in two surfaces) represent intra-layer interaction and layer-layer interaction (inter) is described by the dashed lines. Note that, the connectivity between layer $I$ and layer $II$ is chosen  in such a way that  coupled pairs of nodes have the same index i.e  node $1$ of layer $I$ will be connected to node $1$ of layer $II$ and so on. The layers are controlled by cross adaptive feedback~\cite{Zhang2015} to each other through the inter-layer (dashed lines) interaction. 
\begin{figure}[h] 
\includegraphics[height=!,width=0.48\textwidth]{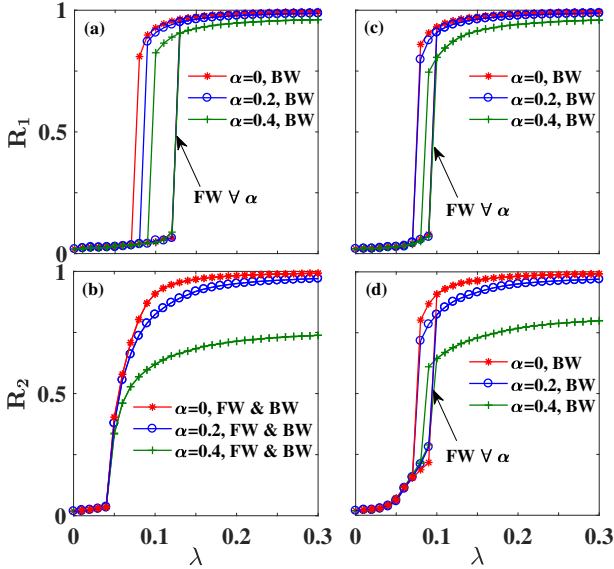}
\caption{Global order parameters of two layers as a function of coupling parameter $\lambda$ for different values of $\alpha$. Synchronization transitions in different layers in absence of inter-layer interactions are shown in figures (a) and (b), while that in presence of inter-layer interactions are shown in (c) and (d) for three different values of $\alpha$.}
\label{simu:rand_sf_N2000_mono_multi}
\end{figure}
The phase $\theta_{i,\sigma(t)}$ of the $i$th oscillator ($i = 1,\dots,N$) of the layer $\sigma (= I,~II)$ is evolved by the differential equation 
\begin{eqnarray}\label{multilayer_system}
\frac{d\theta_{i,\sigma}}{dt} &=& \omega_{i,\sigma} +\lambda{f_{i,\sigma}}\sum_{j=1}^{N} A^ \sigma _{ij}\sin(\theta_{j,\sigma} - \theta_{i,\sigma}-\alpha),
\end{eqnarray}
where $(A_{ij}^{\sigma})_{N\times N}$ and $\omega_{i,\sigma}$ respectively denote the adjacency matrix and natural frequency of the $i$th oscillator of the layer 
$\sigma (= I,~II)$, 
while $\lambda$ and $\alpha$ 
denote the uniform coupling strength and phase-lag parameters  respectively.   The parameter $f_{i,\sigma} = r_{i,\sigma'}~\mathrm{or}~r_{i,\sigma}$ according as there is inter-layer connection or not, where 
$\sigma' = I, II$ and 
$\sigma \neq \sigma'$ i.e. if 
$\sigma = I$, then 
$\sigma' = II$ and vice-versa. The local order parameter $r_{i,\sigma}$ of the layer $\sigma (=I, II)$ is defined by 
$r_{i,\sigma}e^{i\phi_{\sigma}}=(1/k_{i,\sigma})\sum_{j=1}^{N} A^ \sigma _{ij}e^{i\theta_{j,\sigma}}$,
where $k_{i,\sigma}$ and $\phi_{\sigma}$ are the degree of $i$ th node and average local phase of the layer $\sigma (=I, II)$.
Now $R_\sigma e^{i\psi_{\sigma}} = (1/N)\sum_{j=1}^{N}e^{i\theta_{j,\sigma}}$ 
describes the global order parameter of the layer  
$\sigma$ with ${0}\leq{R_\sigma}\leq{1}$ and 
average phase $\psi_\sigma$.

At the outset, we take an Erd\"os-R\'enyi (ER) network of size $N=2000$, mean degree $\langle k\rangle\sim 12$ in layer $I$ and  a heterogeneous   scale-free (SF) network of same size ($N=2000$), scaling exponent $\gamma=2.5$ and mean degree $\langle k\rangle\sim 16$ in second layer ($II$) in absence of inter-layer interaction. In this case, the layers are separated and controlled by their self local order feedback ($f_{i,\sigma} = r_{i,\sigma}, i=1,2,\dots, N$). The natural frequencies of both the layers ($\omega_{i, \sigma}; \sigma = I, II$ ) are drawn from a random homogeneous distribution spread over the range 
$-1$ to $1$.  
We now numerically integrate the system~(\ref{multilayer_system}) using fourth order Runge-Kutta scheme (RK4). For forward (backward) continuation we  adiabatically increase (decrease) the coupling strength 
$\lambda$ with an increment (decrement) of $\delta{\lambda}=0.01$. The adaptively coupled ER network in layer $I$ is found to exhibit ES in presence of weak frustration (see Fig.\ \ref{simu:rand_sf_N2000_mono_multi}(a)).The frustration parameter have no impact on the forward critical point 
$\lambda_f$ where the ensemble transits from incoherence  to coherence during forward continuation. The 
$\lambda_f$ for all $\alpha$ is shown by the black arrow in figure~\ref{simu:rand_sf_N2000_mono_multi}(a). However, during backward continuation, the critical point $\lambda_b$ of transition from coherence to incoherence is found to move towards higher $\lambda$ as $\alpha$ is increased and as a result, the width of hysteresis loop decreases. 
Next the transition is examined in layer $II$ (SF) in the absence of inter-layer interaction. The frustrated dynamics used on top of scale-free network ($\gamma=2.5$ and $\langle k \rangle\sim 14$) does not exhibit ES (see the Fig.\ \ref{simu:rand_sf_N2000_mono_multi}(b)), although the transition looks like hybrid~\cite{Coutinho2013}, or the ES, if exists, is negligibly small \cite{Danziger2016}. The critical transition point for the onset of second order/continuous phase transition is also   independent of $\alpha$, a peculiar but generic behavior is found to occur for all types of adaptive networks.  
\par Now we connect both the layers with cross adaptive inter layer interaction ($f_{i,\sigma} = r_{i,\sigma'}, i=1,2,\dots, N$) 
and we observe that ES is fully established in both the layers. The transition points and hysteresis loops are clearly shown in the Fig.\ \ref{simu:rand_sf_N2000_mono_multi} (c), (d) for layer $I$ and layer $II$ respectively due to the presence of cross adaptive feedback i.e inter layer interaction. 
We also observe that forward critical transition points do not depend on the frustration parameter  present in the systems although it shrinks the ES width if we increase $\alpha$. Note that, the value of order parameter of the second layer $II$ is also slightly increased due to the impact of multiplexity. We now present a detailed and rigorous analytical description of these numerically observed results. We start with an analytical treatment of the monolayer counter parts of the multiplex in absence of inter-layer connections.

\textit{Mean-field analysis}: 
In absence of the inter-layer connections, the layers of the multiplex are separated. Here we perform the mean-field analysis of the isolated layers of the multiplex follwoing the approach proposed in~\cite{Ichinomiya2004}.  Let the density of the nodes at time $t$ with phase $\theta$ for a given degree $k$ and frequency $\omega$ be given by the function $\rho(k,\omega,\theta,t)$ with a proper normalization condition
\begin{eqnarray}
\int_{0}^{2\pi} \rho(k,\omega;\theta,t)d\theta =1.\label{density}
\end{eqnarray}
Avoiding the degree-degree correlation of a network we obtain the probability that a randomly chosen edge is attached to a node with degree $k$ and phase $\theta$ at time $t$ is
\begin{equation}
\frac{kP(k)g(\omega)\rho(k,\omega;\theta,t)}{\int kP(k)dk}.
\end{equation}
Then in the continuum limit, time evolution of the phases of the oscillators of a layer is given by\\
\begin{eqnarray}\label{eqn3}
\frac{d\theta(t)}{dt}=\omega  + \frac{\lambda f k}{\langle k \rangle} \int dk' \int d\omega' \int d\theta' k' P(k') g(\omega')\nonumber \\
\times \rho (k',\omega',\theta',t) \sin(\theta' -\theta -\alpha),
\end{eqnarray}
where $\langle k \rangle = \int kP(k)dk$ 
is the mean degree of the network. 
To maintain the conservation of the oscillators \cite{Kuramoto1984,Ichinomiya2004} in a layer, the density function $\rho$ satisfies the continuity equation
\begin{eqnarray}\label{continuity}
\frac{\partial \rho}{\partial t} + \frac{\partial}{\partial \theta}(\rho v) = 0,
\end{eqnarray}
where $v$ comes from the right hand side of the Eqn.\ (\ref{eqn3}). 

To measure the macroscopic behavior of the oscillators, in the thermodynamic limit we consider the order parameter 
 $R$ given by \cite{Ichinomiya2004} 
\begin{eqnarray}\label{eqn4}
R e^{i\psi}=\frac{1}{\langle k \rangle} \int dk \int d\omega \int d\theta k P(k)g(\omega)\rho(k,\omega,\theta,t) e^{i\theta},
\end{eqnarray}
where $\psi$ 
is the average phase of the oscillators and the value of 
$R$ 
varies in the range 
$0\le R \le 1$. 
Therefore, inserting (\ref{eqn4}),  into (\ref{eqn3}) we obtain
 \begin{eqnarray}
\frac{d\theta}{dt} & = & 
\omega + \lambda r k R \sin(\psi -\theta -\alpha).
\label{eqn5}
\end{eqnarray}
In our present study we consider 
$f_i = r_i (i = 1\dots N)$, where 
$r_i$ 
is the local order parameter. Now we derive the self-consistent equations by setting  the global phase 
$\psi(t) = \Omega t$, where
 $\Omega$ is the group angular frequency. Further  we  introduce a new variable $\phi$ with $\phi(t)=\theta(t)-\psi(t)+\alpha$. In this rotating frame  the equation can be written as 
\begin{eqnarray}
\frac{d\phi}{dt} & = & 
\omega- \Omega - \lambda r k R \sin(\phi), \label{eqn6}.
\end{eqnarray}
 In steady state, we have $\frac{\partial}{\partial t} \rho(k,\omega,\phi) =0$. 
\noindent Therefore, steady state solutions for the density function $\rho$ is given by
\begin{eqnarray}\label{density_function}
 \rho(k,\omega;\phi) &=& \delta \left(\phi-arc\sin{\left(\frac{\omega-\Omega}{\lambda r k R}\right)}\right)~;~~ \abs*{\frac{\omega-\Omega}{\lambda r k R}} \leq 1, \nonumber \\
 &=&\frac{A(k,\omega)}{\abs*{\omega-\Omega -\lambda r k R\sin(\phi)}}~;~~ \abs*{\frac{\omega-\Omega}{\lambda r k R}} > 1, \nonumber
 \end{eqnarray}
 where $\delta$ is the Dirac delta function and $A(k,\omega)$ is the normalization constant.
The first solution corresponds to the synchronous state (a locked version) and second solution is due to desynchronous state (a drift version). Hence the order parameter can be rewritten as 
\begin{eqnarray}\label{r_d_l}
R=\frac{1}{\langle k\rangle} \int \bigg [\int \int_{k_{min}}^{\infty} k P(k) g(\omega) \rho(k,\omega,\phi) \nonumber\\ 
\times e^{i(\phi -\alpha)}{H_1}dk d\omega +\int \int_{k_{min}}^{\infty} k P(k) g(\omega) \nonumber\\
\times \rho(k,\omega,\phi)e^{i(\phi -\alpha)} H_2 dk d\omega \bigg ] d\phi, 
\end{eqnarray}
where $H_1\approx H\left(1-\abs*{\frac{\omega-\Omega}{\lambda r k R}}\right)$ and $H_2\approx H\left(\abs*{\frac{\omega-\Omega}{\lambda r k R}}-1\right)$.

\noindent Here $H$ is heaviside function. Note that, the first part of right hand side of Eqn.\ (\ref{r_d_l}) gives the contribution of locked oscillators  and the second part corresponds to the contribution of drift oscillators  to the order parameter $R$.

Decomposing the Eqn.\ \ref{r_d_l} (see the supplementary for more detailed calculation) we can eventually reach to two coupled equations  
\begin{eqnarray}
R \langle k\rangle&=&\cos \alpha \int \int k P(k) g(\omega) \sqrt{1-\left(\frac{\omega-\Omega}{\lambda r k R} \right)^2} H_1 dk d\omega \nonumber \\
&&+\frac{\sin \alpha}{\lambda r R}(\langle \omega \rangle - \Omega)-\sin \alpha \int \int  k P(k) g(\omega) \nonumber \\
&&\times \frac{\omega-\Omega}{\lambda r k R} \sqrt{1-\left(\frac{\lambda r k R}{\omega-\Omega}\right)^2} H_2 dk d\omega 
\label{real_r}
\end{eqnarray}
and
\begin{eqnarray}
\langle \omega \rangle -\Omega = \lambda r R \tan \alpha \int \int k P(k) g(\omega) ~~~~~~~~~~~~~\nonumber \\
\times \sqrt{1-\left(\frac{\omega-\Omega}{\lambda r k R}\right)^2} H_1 dk d\omega + \int \int P(k) g(\omega) \nonumber \\
\times (\omega-\Omega) \sqrt{1-\left(\frac{\lambda r k R}{\omega-\Omega}\right)^2} H_2 dk d\omega.~~~
\label{im_r}
 \end{eqnarray}
 \noindent
These two coupled equations describe the behavior of  macroscopic order parameter ($R$) and common frequency ($\Omega$) emerged from the coupled network. To solve these coupled equations  (\ref{real_r}-\ref{im_r}) we use  the information  of the network structure (the degree sequences) and the frequency distribution. Note that, in uncorrelated configuration model, the local order parameter behaves similarly as the global order parameter ($r_i \sim R$)  in the thermodynamic limit.  

Next, we extend our analytical approach in multilayer network (Eq.\ (\ref{multilayer_system})) where each layer is controlled by an adaptive feedback from the other layer.
The mathematical formalism leads us to four coupled  equations for two layers in which, first two equations contain the information of two global frequencies of two layers where as the other two equations  contain the information of  order parameters. Based on the equations (\ref{real_r}-\ref{im_r}) we may write (see the details in Supplementary Information \cite{SI})
\begin{eqnarray}\label{im_r_multilayer}
\langle \omega_\sigma \rangle -\Omega_\sigma = \lambda r_{\sigma'} R_\sigma \tan \alpha \int \int k_\sigma P(k_\sigma) g(\omega_\sigma) ~~~~~~~~~~\nonumber \\
\times \sqrt{1-\left(\frac{\omega_\sigma-\Omega_\sigma}{\lambda r_{\sigma'} R_\sigma k_\sigma}\right)^2} {H_1}_\sigma dk_\sigma d\omega_\sigma + \int \int P(k_\sigma) g(\omega_\sigma) \nonumber \\
\times (\omega_\sigma-\Omega_\sigma)\sqrt{1-\left(\frac{\lambda r_{\sigma'} R_\sigma k_\sigma}{\omega_\sigma-\Omega_\sigma}\right)^2} {H_2}_\sigma dk_\sigma d\omega_\sigma ~~~~~
\end{eqnarray}
and 
\begin{eqnarray}\label{real_r_multilayer}
R_\sigma \langle k_\sigma\rangle=\cos \alpha \int \int k_\sigma P(k_\sigma) g(\omega_\sigma) \sqrt{1-\left(\frac{\omega_\sigma-\Omega_\sigma}{\lambda r_{\sigma'} R_\sigma k_\sigma} \right)^2} \nonumber\\
\times {H_1}_\sigma dk_\sigma d\omega_\sigma + \frac{\sin \alpha}{\lambda r_{\sigma'} R_\sigma}(\langle \omega_\sigma \rangle - \Omega_\sigma)-\sin \alpha \int \int k_\sigma \nonumber\\
\times P(k_\sigma)g(\omega_\sigma) \frac{\omega_\sigma-\Omega_\sigma}{\lambda r_{\sigma'} R_\sigma k_\sigma} \sqrt{1-\left(\frac{\lambda r_{\sigma'} R_\sigma k_\sigma}{\omega_\sigma-\Omega_\sigma}\right)^2}\nonumber\\ 
\times{H_2}_\sigma dk_\sigma d\omega_\sigma ~~~~~~~~~~~~ 
\end{eqnarray}
\begin{figure}
\includegraphics[height=!,width=0.48\textwidth]{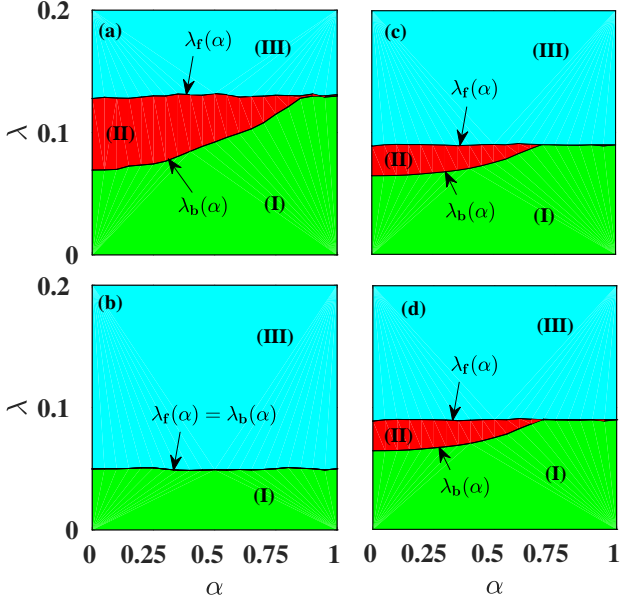}
\caption{Phase diagram on $\alpha-\lambda$ plane showing different regions in two layers of the multiplex in absence ((a) and (b)) and presence ((c) and (d))of inter-layer interactions. Green (I), red (II) and cyan (III) islands respectively represent asynchronous ($R\sim 0$), hysteresis and synchronous ($R\sim 1$) regions.  These regions are separated by solid black line indicating the critical coupling strength for the transition to synchrony during the forward $(\lambda_f)$ and backward $(\lambda_b)$ continuation. }
\label{phase_diagram}
\end{figure}
where, ${H_1}_\sigma \approx H\left(1-\abs*{\frac{\omega_\sigma-\Omega_\sigma}{\lambda r_{\sigma'} R_\sigma k_\sigma}}\right)$ 
 and
  ${H_2}_\sigma \approx H \left(\abs*{\frac{\omega_\sigma-\Omega_\sigma}{\lambda r_{\sigma'} R_\sigma k_\sigma}}-1\right)$ 
  and $\sigma,\sigma'=I,II~(\sigma \neq \sigma')$.
  
Now to validate the self-consistent equations derived above we tune both layers from non-frustrated regime ($\alpha=0$) to highly frustrated regime ($\alpha\sim 1$) in absence as well as in presence of inter-layer interactions. In absence of inter-layer interactions, we perform numerical simulation of the networks in both the layers and demarcate the coherent and incoherent regimes in the $\alpha-\lambda$ plane. Note that, in layers $I$ and $II$ we have considered ER and SF networks of size $N = 2000$ respectively. Figure~\ref{phase_diagram}(a)-(b) shows the impact of phase frustration in the layers $I$ and $II$ in absence of inter-layer interaction. The cyan island (III) represents the coherent/synchronization regime and the green island (I) shows the incoherent regime. The ES regime is shown by the red island. From the figure~\ref{phase_diagram}(a) we observe that ER network exhibits ES for a broad range of frustration ($0<\alpha<0.75$) , where as SF network fails to exhibit ES for any value of $\alpha$ (see Fig.\ \ref{phase_diagram}(b)). Now we consider inter-layer interaction via cross adaptive feedback between the two layers and interestingly, we observe that the multiplexity helps to emerge ES in both the layers as it is evident from the figures~\ref{phase_diagram}(c) and (d). A small red  island  appears in the layer $II$ (SF) signifying the presence of metastable (ES or hysteresis) state which was missing in it's monolayer counterpart. Due to the impact of diffusion, the hysteresis loop (width) is slightly reduced in ER layer.  The black line (upper end of the green island) represents  the backward transition ($\lambda_b$) i.e the changes from coherence state to incoherence state. This numerical boundary has also been validated by self-consistent coupled equations (\ref{real_r}-\ref{im_r}) for mono-layer. The equations (\ref{im_r_multilayer}-\ref{real_r_multilayer}) confirms the analytical validation of backward transition ($\lambda_b$) in multi-layers shown in black line at the upper end of the green island (Fig.\ \ref{phase_diagram}(c)-(d)).  The mean field analysis also confirms the existence of metastable state (ES).  Although the forward transition point (end of metastable state) cannot be predicted from this analysis, a peculiar issue discussed earlier in \cite{Zhang2015,Danziger2016} due to the fluctuation in the order parameter. 
In the next section, we derive exact formula for forward transition to synchronization analytically.\\   

{\it Forward transition.}  To determine the exact forward transition point ($\lambda_f$) analytically, we proceed by perturbing the  completely incoherent state. In completely incoherent state $\rho(k,\theta,\omega,t)=\frac{1}{2\pi}$ and it is perturbed with small amplitude $\eta$ as 
\begin{eqnarray}\label{density_perturbation}
\rho(k,\theta,\omega,t)=\frac{1}{2\pi}+\epsilon\eta(k,\theta,\omega,t),~\mathrm{where}~\epsilon \ll 1.
\end{eqnarray}
Since $\int_{o}^{2\pi} \eta(k,\theta,\omega,t) d\theta = 0$, we have $R' e^{i\psi}= \epsilon R e^{i\psi}$ which imply $R'= \epsilon R$ (See supplementary material) and 
\begin{eqnarray}\label{eta_R}
R e^{i\psi} = \frac{1}{\langle k \rangle} \int \int \int k P(k)g(\omega)\eta(k,\omega,\theta,t) e^{i\theta} d\theta d\omega dk .
\end{eqnarray}
The flow velocity function $v(t)$ is defined by,\\
\begin{eqnarray}\label{velocity_func}
v(t)=\frac{d\theta}{dt}&=&\omega(t)+\lambda r k R' \sin(\psi-\theta-\alpha) \nonumber \\
&=& \omega(t)+\epsilon \lambda r k R \sin(\psi-\theta-\alpha).
\end{eqnarray}
By substituting (\ref{density_perturbation}),(\ref{eta_R}),(\ref{velocity_func}) into (\ref{continuity}) and neglecting the higher order term of $\epsilon$, we have\\
\begin{eqnarray}\label{partial_eta}
\frac{\partial \eta}{\partial t}=-\omega \frac{\partial \eta}{\partial \theta}+\frac{\lambda r k R \cos(\psi-\theta-\alpha)}{2\pi}.
\end{eqnarray}
Now complex Fourier series of the function $\eta(k,\theta,\omega,t)$ is given by,
\begin{eqnarray}\label{eta_fourier}
\eta(k,\theta,\omega,t)=c(k,\omega,t)e^{i\theta}+c^*(k,\omega,t)e^{-i\theta}+\nonumber\\ \eta^\bot(k,\theta,\omega,t),
\end{eqnarray}
where $\eta^\bot$ represents the higher order Fourier harmonic terms. 
Eventually we may write (for detailed calculations, see supplementary \cite{SI})
\begin{eqnarray}\label{rcos}
&&R\cos(\psi-\theta-\alpha) ~~~~~~~~~~~~~~\nonumber \\
&=&\frac{\pi}{\langle k \rangle}[e^{-i(\theta+\alpha)}\int \int k P(k)g(\omega) c^*(k,\omega,t) d\omega dk~~~~~ \nonumber \\
&&+e^{i(\theta+\alpha)}\int \int k P(k)g(\omega) c(k,\omega,t) d\omega dk]
\end{eqnarray}
Using Eqn.\ (\ref{eta_fourier}), (\ref{rcos}) into the Eqn.\ (\ref{partial_eta}) and comparing the coefficients of $e^{i\theta}$ we arrive at
\begin{eqnarray}\label{c_eqn}
\frac{\partial c(k,\omega,t)}{\partial t}=-i \omega c(k,\omega,t)+\frac{\lambda r k}{2 \langle k \rangle}e^{i \alpha}\int_{k_{min}}^{\infty} \int k' P(k') \nonumber \\
\times g(\omega') c(k',\omega',t) d\omega' dk'.~~~~~~~~~~~~~~
\end{eqnarray}
We now assume $c(k,\omega,t)=A(k)B(\omega)e^{\mu t}$ and substituting it into the Eqn.\ \ref{c_eqn} we get
\begin{eqnarray}\label{putting_c}
\mu A(k)B(\omega)=-i \omega A(k)B(\omega)+\frac{\lambda r k}{2 \langle k \rangle}e^{i \alpha}\int_{k_{min}}^{\infty} \int_{-\infty}^{\infty} k' \nonumber \\
\times P(k')g(\omega')A(k')B(\omega') d\omega' dk'.~~~~~~~~~~~~~~
\end{eqnarray}
Finally, the equation looks like
\begin{eqnarray}\label{forward_main}
1=\frac{\lambda r \langle k^2 \rangle}{2 \langle k \rangle}e^{i \alpha}\int \frac{(\mu-i\omega)g(\omega)}{\mu^2 + \omega^2}d\omega.
\end{eqnarray}
For symmetric uniform frequency distribution within an interval 
$[-a,a]$,  $g(\omega)= \frac{1}{2a}$ for $-a\leq\omega\leq a$ and $g(\omega)=0$, otherwise. In that case, from (\ref{forward_main}), we get
\begin{eqnarray}
 \lambda_f=\frac{4a\langle k \rangle \cos\alpha}{\pi R \langle k^2 \rangle},
\label{lambda_f}
\end{eqnarray}
by taking  the limit $\mu \rightarrow 0$ for marginal stability and  
$r\sim R$ 
due to the annealed network approximation.  Here 
$\lambda_f$ 
gives the forward transition point where the incoherent state looses it's stability. The expression of $\lambda_f$ (\ref{lambda_f}) 
combines the network structure ($\langle k \rangle $  and  $\langle k^2 \rangle$), with system parameters: frequency distribution 
$g(\omega)$ 
and phase-frustration parameter $\alpha$. 
 Note that, the  ratio of the first non-zero order parameter ($R$) and 
 $\cos \alpha$ 
 remains constant for a given network (numerical verification is not shown here), therefore 
 $\lambda_f$ 
 depends only on network structure and frequency distribution.  The approach fits with our numerical result shown in the Fig.\ \ref{simu:rand_sf_N2000_mono_multi} (a)-(b) where forward point does not depend on 
 $\alpha$. The same behavior is  also shown with black lines (lower end of cyan island) in   Fig.\ (\ref{phase_diagram} (a)-(b)). As expected, the black lines  are almost parallel to $x$-axis ($\alpha$).  In the absence of inter-layer connection we calculate the forward critical point from (\ref{lambda_f}), in which the values  for the layer $I$ and layer $II$ are
 $\lambda_f\sim 0.12$ and $\lambda_f\sim 0.04$ respectively. We have verified our result for several choices of initial states. Note that, the forward transition point of scale-free layer (II) is close to zero, as we consider $\gamma=2.5$, where $\frac{<k>}{<k^2>}$ converges to zero for $N\rightarrow \infty$. Due to finite network size we get a small nonzero $\lambda_f$ value.
Next we have extended our forward transition calculation in multilayer network in presence of inter-layer interaction and obtain 
\begin{eqnarray}
\lambda_{\sigma_c}=\frac{4a\langle k_\sigma \rangle \cos\alpha}{\pi r_{\sigma'} \langle k_\sigma^2 \rangle};~~
\sigma,\sigma'=I,II~(\sigma \neq \sigma')
\label{lambda_f_multi}.
\end{eqnarray}
Due to the diffusive interaction, critical points of both layers remain close to each other, the total hysteresis area of both layers (shown in red color in Fig.\ref{phase_diagram} (c)-(d)) remain identical which is also confirmed by our analytical 
expression (\ref{lambda_f_multi}).  We would like to mention that, considering  SF
 network in lower $\gamma$ ($\gamma<3$), the theoretical forward point comes closer to the backward point, although a small width of ES is confirmed for different set of realizations. 

\begin{figure}[h]
\includegraphics[height=!,width=0.48\textwidth]{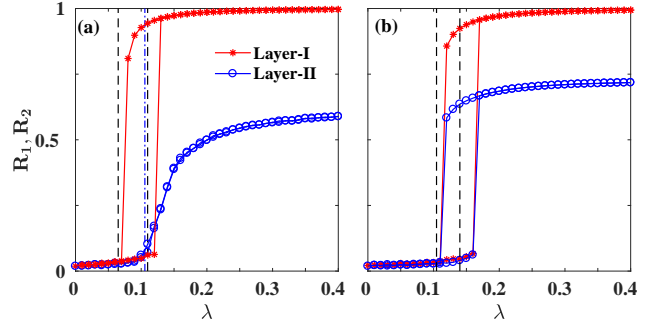}
\caption{Order parameters of two layers are shown as a function of $\lambda$ in absence (a) and presence (b) of inter-layer interactions. Vertical dashed lines represent critical points determined from the self consistent equations (\ref{real_r}-\ref{im_r}),  (\ref{im_r_multilayer}-\ref{real_r_multilayer}) and from the Eqn.\ \ref{lambda_f}.
}
\label{multilayer_alpha_vary}
\end{figure}
Now we will show the impact of multiplexity over frustration. We consider a multiplex network in which ER networks are used on both the layers in absence of inter-layer interactions. To realize ES, we consider non-frustrated environment in layer $I$ where $\alpha=0$. The changes of order parameter (with respect to $\lambda$) is shown in the Fig.\ref{multilayer_alpha_vary} (a) with red line and the blue line shows a continuous transition in layer $II$ due to the presence of strong frustration ($\alpha=1$). 
Surprisingly,  as the inter-layer interactions are switched on, both the layers exhibit ES behavior as shown in  Fig.\  \ref{multilayer_alpha_vary} (b). Note that, the value  of the order parameter of layer $II$ (blue color) is enhanced in multilayer due to the impact of layer $I$, an interesting phenomena never discussed before in phase frustrated environment. The dashed vertical lines represent the critical coupling strengths calculated analytically.  The Backward points are close agreement with the numerical values, however the forward points have small discrepancy (mainly in multiplex structure) with the numerics. Although, the multiplexity wins over frustration by setting all layers to ES. We have also tested and verified this feature for SF networks with two different frustration terms ($\alpha=0, \alpha=1$) (See the Supplementary section \cite{SI}). 

We found here that, a network can impact or affect the sharp transition of a frustrated system. For instance, a SF network in lower $\gamma$ can not undergo first order transition. On the other hand, for a given network, a strong phase frustration inhibits ES by promoting  a continuous transition to synchronization. We have established that, we can transform a continuous transition to a discontinuous transition by feeding them into multiplex network in which each layer is adaptively controlled by the other layer. We have numerically shown the emergence of ES and our analytical approach predicts the accurate condition of emergence of ES in multilayer as well as in its monolayer counter part. Using the mean field analysis and a perturbative approach around incoherent state, we calculate the critical coupling strengths for the onset of synchronization both for forward and backward transition. Our approach can easily be extended to other frequency distributions. \\

\textit{Acknowledgements}:
PK acknowledges support from  DST, India under the DST-INSPIRE scheme (Code: IF140880). CH is supported by INSPIRE faculty award scheme.

\pagebreak
\begin{widetext}
\section{Supplimentary Information}
\subsection{Mean-field analysis}
We start with a  network of 
$N$ Sakaguchi-Kuramoto like phase oscillators where the phase 
 $\theta_i(t) (i = 1\dots N)$ 
 of each oscillator is driven by the dynamic equation
\begin{eqnarray}\label{eqn1}
\frac{d\theta_i}{dt} &=& \omega_i +\lambda{f_i}\sum_{j=1}^{N} A_{ij}\sin(\theta_j - \theta_i-\alpha),
\end{eqnarray}
where, $\omega_i$ represents the natural frequency of the 
$i^{th}$ 
oscillator for 
$i = 1, \dots ,N$; $A_{ij}$ 
is the $ij^{th}$ 
element of the adjacency matrix 
$A = (A_{ij})_{N\times N}$ 
such that 
$A_{ij} = 1$ if $i^{th}$ and $j^{th}$ oscillators are connected and $A_{ij} = 0$ otherwise, 
$\alpha$ 
is the phase-lag parameter whose value lies in the range 
$0\leq \alpha < \frac{\pi}{2}$ 
and $\lambda$ is the coupling strength.
The considered adaptive coupling  $f_i$ of each node  is adopted from   the order parameter ($r_i$) emerging from the adjacent nodes of node $i$.
 The local order parameter for the $i$th oscillator is defined as
\begin{eqnarray}\label{local_order_parameter}
r_i(t)e^{i\phi}=(1/k_i)\sum_{j=1}^{N} A_{ij}e^{i\theta_j},
\end{eqnarray}
where, ${0}\leq{r_i}\leq{1}$ and $\phi$ denotes the phase averaged over the ensemble of neighbors. \\
Here we set the frequencies $\omega_i$  from a random homogeneous distribution in the range $-1$ to $1$. We now numerically simulate the adaptive Sakaguchi-Kuramoto model both on ER and SF network with $N=1000$ for different values of $\alpha$. For numerical simulation, we are using fourth-order Runga-Kutta method with time step $\delta{t}=0.01$ to integrate the system $(1)$. The level of synchronization in the network can be measured by the global order parameter $R$ which is defined by
\begin{eqnarray}\label{Global}
Re^{i\psi} = (1/N)\sum_{j=1}^{N}e^{i\theta_j},
\end{eqnarray}
where, ${0}\leq{R}\leq{1}$ and 
$\psi$ denotes the average phase of the network.\\
In our next step we want to show the general scenario for the case of multilayer networks with different topological features. In this purpose we have taken two independent networks (I and II) with the same size $N$. Note
that, due to the presence of multiplexity, the connectivity
between layer ${\it I}$ and layer ${\it II}$ is chosen such a way that coupled pairs of nodes have the same index i.e node $1$ of layer ${\it I}$ will be connected to node $1$ of layer II and so on and so forth. The equations of motion can be written as
\begin{eqnarray}\label{multilayer_system}
\frac{d\theta_{i,1}}{dt} &=& \omega_{i,1} +\lambda{f_{i,1}}\sum_{j=1}^{N} A^1_{ij}\sin(\theta_{j,1} - \theta_{i,1}-\alpha), \nonumber \\
\frac{d\theta_{i,2}}{dt} &=& \omega_{i,2} +\lambda{f_{i,2}}\sum_{j=1}^{N} A^2_{ij}\sin(\theta_{j,2} - \theta_{i,2}-\alpha),
\end{eqnarray}
where, i = 1,\dots,N and the subscripts $1$ and $2$ stands for the layers I and II respectively. In equation (4) and (5) the average degree is $<k_1>=(1/N)\sum_{i=1}^{N}{k_{i,1}}$ and $<k_2>=(1/N)\sum_{i=1}^{N}{k_{i,2}}$ for the $1^{st}$ layer and the $2^{nd}$ layer. The parameters $f_{i,1}$ and $f_{i,2}$ account for the coupling between two layers. Precisely we set $f_{i,1}=r_{i,2}$ and $f_{i,2}=r_{i,1}$, where $r_{i,1}$ and $r_{i,2}$ are defined by,\\
\begin{eqnarray}\label{multilayer_local}
r_{i,1}e^{i\phi_{1}}=(1/k_{i,1})\sum_{j=1}^{N} A^1_{ij}e^{i\theta_{j,1}},\\
r_{i,2}e^{i\phi_{2}}=(1/k_{i,2})\sum_{j=1}^{N} A^2_{ij}e^{i\theta_{j,2}},
\end{eqnarray}
In other words, all the oscillators in layer ${\it I}$ is here adaptively controlled by the local order parameters of the corresponding nodes on layer ${\it II}$, and vice versa.
Now let the global order parameters of layer ${\it I}$ and layer ${\it II}$ are respectively $R_1$ and $R_2$, which are defined by,

\begin{eqnarray}\label{multilayer_global}
R_1e^{i\psi_{1}} = (1/N)\sum_{j=1}^{N}e^{i\theta_{j,1}},\\
R_2e^{i\psi_{2}} = (1/N)\sum_{j=1}^{N}e^{i\theta_{j,2}},
\end{eqnarray}
where ${0}\leq{R_1,R_2}\leq{1}$ and $\psi_1$, $\psi_2$ denotes the average network's phase for 1st layer and 2nd layer respectively.

In the following we will move with the mean-field approach proposed in \cite{Ichinomiya2004, Kundu2017}, let the density of the nodes at time 
$t$ with phase 
$\theta$ 
for a given degree 
$k$ and frequency 
$\omega$ be given by the function $\rho(k,\omega,\theta,t)$, and we can normalize it as  
\begin{eqnarray}
\int_{0}^{2\pi} \rho(k,\omega;\theta,t)d\theta =1.\label{density}
\end{eqnarray}
We assume that the nodes of the network are not degree correlated and therefore the probability that a randomly chosen edge is attached to a node with degree $k$ and phase $\theta$ at time $t$ can be written as 
\begin{equation}
\frac{kP(k)g(\omega)\rho(k,\omega;\theta,t)}{\int kP(k)dk}.
\end{equation}
Then with the continuum limit, Eqn.\ (\ref{eqn1}) can be written as\\
\begin{eqnarray}\label{eqn3}
\frac{d\theta(t)}{dt}=\omega  + \frac{\lambda f k}{\langle k \rangle} \int dk' \int d\omega' \int d\theta' k' P(k') g(\omega') \rho (k',\omega',\theta',t) \sin(\theta' -\theta -\alpha),
\end{eqnarray}
where $\langle k \rangle = \int kP(k)dk$ denotes the mean degree of the network. 
Now for the conservation of the oscillators for Eqn.\ (\ref{eqn1}), the density function $\rho$ satisfies the continuity equation
\begin{eqnarray}\label{continuity}
\frac{\partial \rho}{\partial t} + \frac{\partial}{\partial \theta}(\rho v) = 0
\end{eqnarray}
where $v$ is the right hand side of the Eqn.\ (\ref{eqn3}). 

To measure the macroscopic behavior of the oscillators, in the thermodynamic limit we consider the order parameter $R$ given by \cite{Ichinomiya2004} 
\begin{eqnarray}\label{eqn4}
R e^{i\psi}&=&\frac{1}{\langle k \rangle} \int dk \int d\omega \int d\theta k P(k)g(\omega)\rho(k,\omega,\theta,t) e^{i\theta}, 
\end{eqnarray}
where $\psi$ is the average phase of the oscillators and the value of $R$ varies in the range $0\le R \le 1$. Therefore,  equation~(\ref{eqn3}) can be written as
 \begin{eqnarray}
\frac{d\theta}{dt} & = & 
\omega + \lambda r k R \sin(\psi -\theta -\alpha).
\label{eqn5}
\end{eqnarray}
In our present study we consider 
$f_i = r_i (i = 1\dots N)$, where 
$r_i$ 
is the local order parameter (\ref{local_order_parameter}). Now we set the global phase 
$\psi(t) = \Omega t$ 
to derive the self-consistent equations where 
$\Omega$ is the group angular frequency and introduce a new variable $\phi$ with $\phi(t)=\theta(t)-\psi(t)+\alpha$. 
In terms of this new variable,  Eqn.\ (\ref{eqn5}) can be written as 
\begin{eqnarray}
\frac{d\phi}{dt} & = & 
\omega- \Omega - \lambda r k R \sin(\phi).\label{eqn6}
\end{eqnarray}
\noindent
\begin{figure}
\includegraphics[height=!,width=0.55\textwidth]{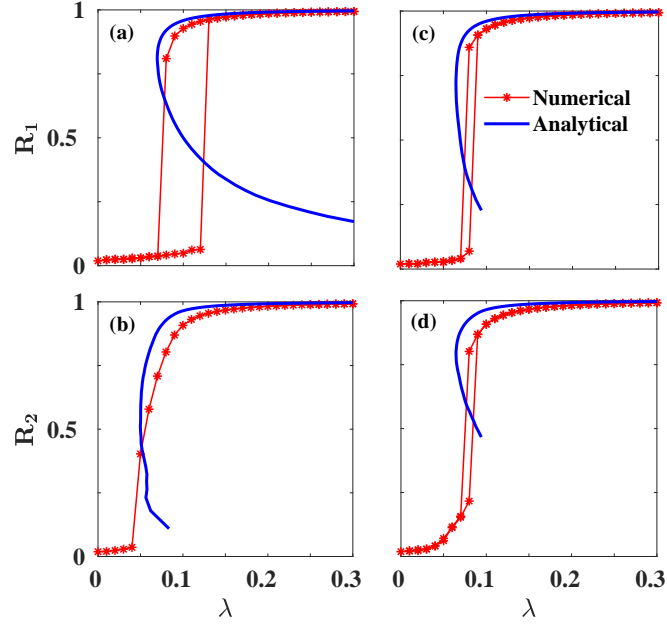}
\caption{Comparison of synchronization transition of two layers for the numerical result with the analytical one. Figure (a) and (c) denote the 1st layer equipped with ER network of size $N=2000$ and $\langle k \rangle=12$ and figure (b) and (d) indicate the synchronization diagram of 2nd layer equipped with scale-free network of size $N=2000$, $\langle k \rangle=16$ and $\gamma=2.5$. Every simulation computed with the phase frustration value $\alpha=0$. The red line indicates the numerical simulation and blue line indicates the analytical simulation computed with the equations (\ref{omega_x}),(\ref{r_x_lambda}),(\ref{r_x}) in Figure (a) and (b) and with the equations (\ref{im_r_multilayer}),(\ref{real_r_multilayer}) in Figure (c) and (d). 1st column of the figure shows the synchronization phenomenon when the two layers are not connected to each other and 2nd column shows the same when the layers are connected through adaptive links like the schematic diagram (Fig.(1) of main text). }
\label{simu:rand_sf_N2000_mono_multi}
\end{figure}\\
\noindent
The equation of continuity (\ref{continuity}) then takes the form
 \begin{eqnarray}
 \frac{\partial}{\partial t} \rho(k,\omega,\phi) + \frac{\partial}{\partial \phi} [v_\phi \rho(k,\omega,\phi)]=0, \label{eqn_cont}
 \end{eqnarray}
where $v_\phi = \frac{d\phi}{dt}$. In steady state, we have $\frac{\partial}{\partial t} \rho(k,\omega,\phi) =0$. 
\noindent Therefore, steady state solutions for the density function $\rho$ is given by
\begin{eqnarray}\label{density_function}
 \rho(k,\omega;\phi) &=& \delta \left(\phi-arc\sin{\left(\frac{\omega-\Omega}{\lambda r k R}\right)}\right)~;~~ \abs*{\frac{\omega-\Omega}{\lambda r k R}} \leq 1, \nonumber \\
 &=&\frac{A(k,\omega)}{\abs*{\omega-\Omega -\lambda r k R\sin(\phi)}}~;~~ \abs*{\frac{\omega-\Omega}{\lambda r k R}} > 1
 \end{eqnarray}
 where $\delta$ is the Dirac delta function and $A(k,\omega)$ is the normalization constant.

\begin{figure}
\includegraphics[height=3in,width=!]{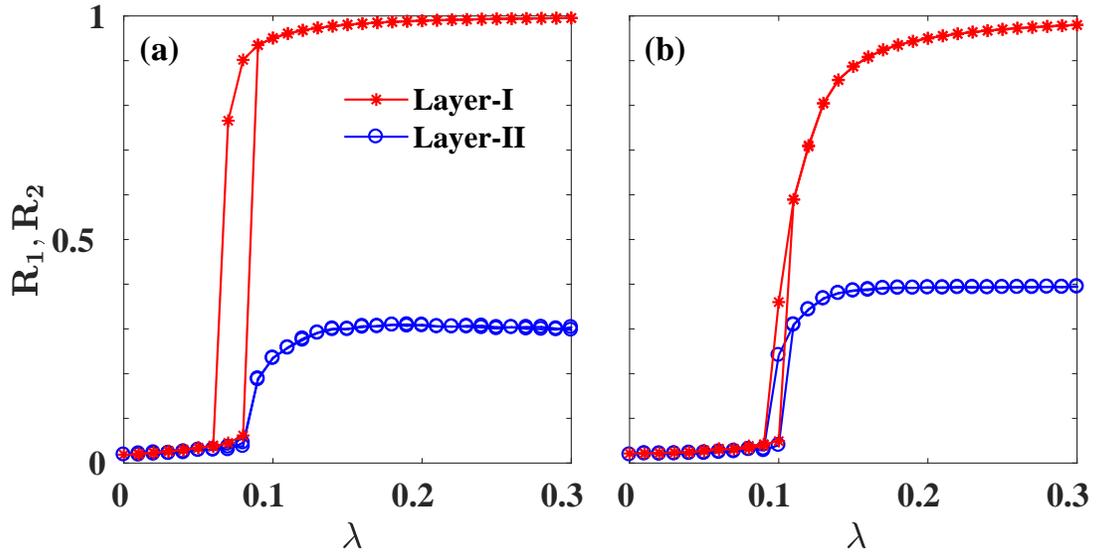}
\caption{Figure (a) indicates the order parameter evaluation as a function of $\lambda$ in the absence of inter layer interaction. The red line shows hysteresis for $\alpha=0$ in layer $I$ and the blue line shows synchronization pattern of layer $II$ when $\alpha=1$. There is no ES in layer $II$. (b) Both of the network reveals ES (shown by red and blue line) in the presence of multiplexity in which, two layers are connected through  adaptive links. Both the layers in two networks equipped with a scale free network of size $N=2000$, $\langle k \rangle=14$ and, $\gamma=3.5$.}
\label{simu:sf_sf_N2000_alpha0_1}
\end{figure}
 
The first solution corresponds to the synchronous state and second solution is due to desynchronous state. Hence the order parameter can be rewritten as 
\begin{eqnarray}
R &=& \frac{1}{\langle k\rangle} \int \bigg [\int \int_{k_{min}}^{\infty} k P(k) g(\omega) \rho(k,\omega,\phi) e^{i(\phi -\alpha)} H\left(1-\abs*{\frac{\omega-\Omega}{\lambda r k R}}\right)dk d\omega \nonumber \\
 &+& \int \int_{k_{min}}^{\infty} k P(k) g(\omega)\rho(k,\omega,\phi) e^{i(\phi -\alpha)}H\left(\abs*{\frac{\omega-\Omega}{\lambda r k R}}-1\right) dk d\omega \bigg ] d\phi, \label{r_d_l}
\end{eqnarray}
\noindent where $H$ is heaviside functions.Here the first part of right hand side of Eq.\ (\ref{r_d_l}) gives the contribution of locked oscillators  and the second part denotes the contribution of drift oscillators  to the order parameter $r$.

Hence the contribution of locked oscillators to the order parameter is
 \begin{eqnarray}
R_{l}=\left[\frac{\cos \alpha}{\langle k\rangle}\int_{k_{min}}^{\infty} \int k P(k) g(\omega) \sqrt{1-\left(\frac{\omega-\Omega}{\lambda r k R}\right)^2} dk d\omega +  \frac{\sin \alpha}{\langle k\rangle}\int_{k_{min}}^{\infty} \int k P(k) g(\omega) \frac{\omega-\Omega}{\lambda r k R} dk d\omega \right]H\left(1-\abs*{\frac{\omega-\Omega}{\lambda r k R}}\right) \nonumber \\
-i\left[ \frac{\sin \alpha}{\langle k\rangle}\int_{k_{min}}^{\infty} \int k P(k) g(\omega) \sqrt{1-\left(\frac{\omega-\Omega}{\lambda r k R}\right)^2} dk d\omega -\frac{\cos \alpha}{\langle k\rangle}\int_{k_{min}}^{\infty} \int k P(k) g(\omega) \frac{\omega-\Omega}{\lambda r k R} dk d\omega \right]H\left(1-\abs*{\frac{\omega-\Omega}{\lambda r k R}}\right) \nonumber \\
\label{r_lock}
 \end{eqnarray}
 Now the contribution of drift oscillators to the order parameter is given by
 \begin{eqnarray}
R_{d}=\frac{(\sin \alpha +i\cos \alpha)}{\langle k\rangle}\int_{k_{min}}^{\infty} \int P(k) g(\omega)\frac{\omega-\Omega}{\lambda r R} \left[1-\sqrt{1-\left(\frac{\lambda r k R}{\omega-\Omega}\right)^2}  \right]dk d\omega H\left(\abs*{\frac{\omega-\Omega}{\lambda r k R}}-1\right)
\label{r_drift}
 \end{eqnarray}
 
 Hence we get 
 $R=R_l +R_d$, where 
 $R_l$ and $R_d$ 
 are given by Eq.\ (\ref{r_lock}) and Eq.\ (\ref{r_drift}) respectively.
Now comparing the real and imaginary parts we get
\begin{eqnarray}
R \langle k\rangle= \cos \alpha \int_{k_{min}}^{\infty} \int k P(k) g(\omega) \sqrt{1-\left(\frac{\omega-\Omega}{\lambda r k R} \right)^2} H\left(1-\abs*{\frac{\omega-\Omega}{\lambda r k R}}\right) dk d\omega + \frac{\sin \alpha}{\lambda r R}(\langle \omega \rangle - \Omega) \nonumber \\-\sin \alpha \int_{k_{min}}^{\infty} \int  k P(k) g(\omega)\frac{\omega-\Omega}{\lambda r k R} \sqrt{1-\left(\frac{\lambda r k R}{\omega-\Omega}\right)^2} H\left(\abs*{\frac{\omega-\Omega}{\lambda r k R}}-1\right) dk d\omega 
\label{real_r}
\end{eqnarray}
and
\begin{eqnarray}
\langle \omega \rangle -\Omega = \lambda r R \tan \alpha \int_{k_{min}}^{\infty} \int k P(k) g(\omega) \sqrt{1-\left(\frac{\omega-\Omega}{\lambda r k R}\right)^2} H\left(1-\abs*{\frac{\omega-\Omega}{\lambda r k R}}\right) dk d\omega \nonumber \\
+\int_{k_{min}}^{\infty} \int P(k) g(\omega)(\omega-\Omega) \sqrt{1-\left(\frac{\lambda r k R}{\omega-\Omega}\right)^2} H \left(\abs*{\frac{\omega-\Omega}{\lambda r k R}}-1\right) dk d\omega
\label{im_r}
 \end{eqnarray}
  In the mean-field framework we can take $r_i=R$.  
 Now to simplify our calculation we introduce a variable 
 $x=\lambda R^2$ and substituting in Eqn.\ (\ref{real_r}) and Eqn.\ (\ref{im_r}) we obtain three set of equations for the unknown parameters  $\Omega$,$R$ and $x$. Here it is convenient to consider 
 $\Omega$ and $R$ as the functions of 
 $x$

 \begin{eqnarray}
\langle \omega \rangle -\Omega(x) =x \tan \alpha \int_{k_{min}}^{\infty} \int k P(k) g(\omega) \sqrt{1-\left(\frac{\omega-\Omega(x)}{x k}\right)^2} H\left(1-\abs*{\frac{\omega-\Omega}{x k}}\right) dk d\omega \nonumber \\
\int_{k_{min}}^{\infty} \int p(k) g(\omega) (\omega-\Omega(x)) \sqrt{1-\left(\frac{x k}{\omega-\Omega(x)}\right)^2} H\left(\abs*{\frac{\omega-\Omega}{x k}}-1\right) dk d\omega,
 \label{omega_x}
 \end{eqnarray}

 \begin{eqnarray}
 R(x)=\sqrt{\frac{x}{\lambda}}
 \label{r_x_lambda},
 \end{eqnarray}
 and
 \begin{eqnarray}
R(x)\langle k\rangle=\cos \alpha \int_{k_{min}}^{\infty} \int k P(k) g(\omega) \sqrt{1-\left(\frac{\omega-\Omega}{x k}\right)^2} H\left(1-\abs*{\frac{\omega-\Omega}{x k}}\right) dk d\omega + \frac{\sin \alpha}{x}(\langle \omega \rangle- \Omega) \nonumber \\-\sin \alpha \int_{k_{min}}^{\infty} \int k P(k) g(\omega) \frac{\omega-\Omega}{x k} \sqrt{1-\left(\frac{x k}{\omega-\Omega}\right)^2} H\left(\abs*{\frac{\omega-\Omega}{x}}-1\right) dk d\omega
\label{r_x}.
\end{eqnarray}
Solving Eq.(\ref{omega_x}) and Eq.(\ref{r_x_lambda}) we find $x$ and the group angular frequency $\Omega$. Inserting them into Eq.\ (\ref{r_x}) we can find the order parameter $R$. 

We repeat here the same approximation for bi-layer multiplex network. Taking the  Eq.\ (\ref{multilayer_system}) into account, we get two set of equations for two layers. Where, from the first set of equations we can get the critical values of frequencies and from the second set we can get  order parameters of two layers.

\begin{eqnarray}
\langle \omega_1 \rangle -\Omega_1 = \lambda R_2 R_1 \tan \alpha \int_{k_{min}}^{\infty} \int k_1 P(k_1) g(\omega_1) \sqrt{1-\left(\frac{\omega_1-\Omega_1}{\lambda R_2 R_1 k_1}\right)^2} H\left(1-\abs*{\frac{\omega_1-\Omega_1}{\lambda R_2 R_1 k_1}}\right) dk_1 d\omega_1 \nonumber \\
+\int_{k_{min}}^{\infty} \int P(k_1) g(\omega_1)(\omega_1-\Omega_1) \sqrt{1-\left(\frac{\lambda R_2 R_1 k_1}{\omega_1-\Omega_1}\right)^2} H \left(\abs*{\frac{\omega_1-\Omega_1}{\lambda R_2 R_1 k_1}}-1\right) dk_1 d\omega_1 \nonumber \\
\langle \omega_2 \rangle -\Omega_2 = \lambda R_1 R_2 \tan \alpha \int_{k_{min}}^{\infty} \int k_2 P(k_2) g(\omega_2) \sqrt{1-\left(\frac{\omega_2-\Omega_2}{\lambda R_1 R_2 k_2}\right)^2} H\left(1-\abs*{\frac{\omega_2-\Omega_2}{\lambda R_1 R_2 k_2}}\right) dk_2 d\omega_2 \nonumber \\
+\int_{k_{min}}^{\infty} \int P(k_2) g(\omega_2)(\omega_2-\Omega_2) \sqrt{1-\left(\frac{\lambda R_1 R_2 k_2}{\omega_2-\Omega_2}\right)^2} H \left(\abs*{\frac{\omega_2-\Omega_2}{\lambda R_1 R_2 k_2}}-1\right) dk_2 d\omega_2
\label{im_r_multilayer}
\end{eqnarray}
Now by considering the real part,
\begin{eqnarray}
R_1 \langle k_1\rangle= \cos \alpha \int_{k_{min}}^{\infty} \int k_1 P(k_1) g(\omega_1) \sqrt{1-\left(\frac{\omega_1-\Omega_1}{\lambda R_2 R_1 k_1} \right)^2} H\left(1-\abs*{\frac{\omega_1-\Omega_1}{\lambda R_2 R_1 k_1}}\right) dk_1 d\omega_1 + \frac{\sin \alpha}{\lambda R_2 R_1}(\langle \omega_1 \rangle - \Omega_1) \nonumber \\
-\sin \alpha \int_{k_{min}}^{\infty} \int  k_1 P(k_1) g(\omega_1)\frac{\omega_1-\Omega_1}{\lambda R_2 R_1 k_1} \sqrt{1-\left(\frac{\lambda R_2 R_1 k_1}{\omega_1-\Omega_1}\right)^2} H\left(\abs*{\frac{\omega_1-\Omega_1}{\lambda R_2 R_1 k_1}}-1 \right) dk_1 d\omega_1 \nonumber\\
R_2 \langle k_2\rangle= \cos \alpha \int_{k_{min}}^{\infty} \int k_2 P(k_2) g(\omega_2) \sqrt{1-\left(\frac{\omega_2-\Omega_2}{\lambda R_1 R_2 k_2} \right)^2} H\left(1-\abs*{\frac{\omega_2-\Omega_2}{\lambda R_1 R_2 k_2}}\right) dk_2 d\omega_2 + \frac{\sin \alpha}{\lambda R_1 R_2}(\langle \omega_2 \rangle - \Omega_2) \nonumber \\
-\sin \alpha \int_{k_{min}}^{\infty} \int  k_2 P(k_2) g(\omega_2)\frac{\omega_2-\Omega_2}{\lambda R_1 R_2 k_2} \sqrt{1-\left(\frac{\lambda R_1 R_2 k_2}{\omega_2-\Omega_2}\right)^2} H\left(\abs*{\frac{\omega_2-\Omega_2}{\lambda R_1 R_2 k_2}}-1 \right) dk_2 d\omega_2
\label{real_r_multilayer}
\end{eqnarray}
In Fig.\ (\ref{simu:rand_sf_N2000_mono_multi}) we have taken ER network of size $N=2000$ and $\langle k \rangle=12$ in $1^{st}$ layer and scale-free network of size $N=2000$, $\langle k \rangle=16$ and $\gamma=2.5$ in $2^{nd}$ layer. We have done numerical simulations with the phase frustration value $\alpha=0$ using fourth order RK method with step size $0.01$. The red line indicates the numerical simulation and blue line indicates the analytical simulation computed with the equations (\ref{omega_x}),(\ref{r_x_lambda}),(\ref{r_x}) in Figure (a) and (b) and with the equations (\ref{im_r_multilayer}),(\ref{real_r_multilayer}) in Figure (c) and (d). 1st column of the figure shows the synchronization phenomenon in absence of inter layer interaction and 2nd column shows the same when the layers are connected through adaptive links  (Fig.\ (1) of main text).
\subsection{Perturbation analysis for Forward transition}
Now to better understand the transition point of synchronization (forward critical point) we will check the stability of the incoherent state. For this purpose we take a small perturbation around the completely incoherent state: $\rho_0(k,\theta,\omega,t)=\frac{1}{2\pi}$ with $\epsilon \ll 1$:
\begin{eqnarray}\label{density_perturbation}
\rho(k,\theta,\omega,t)=\frac{1}{2\pi}+\epsilon\eta(k,\theta,\omega,t)
\end{eqnarray}
Since $\int_{o}^{2\pi} \eta(k,\theta,\omega,t) d\theta = 0$, we have
\begin{eqnarray}
R' e^{i\psi} &=& \frac{1}{\langle k \rangle} \int \int \int k P(k)g(\omega)\rho(k,\omega,\theta,t) e^{i\theta} d\theta d\omega dk \nonumber \\
&=& \frac{\epsilon}{\langle k \rangle} \int \int \int k P(k)g(\omega)\eta(k,\omega,\theta,t) e^{i\theta} d\theta d\omega dk= \epsilon R e^{i\psi}
\end{eqnarray}
We now get $R'= \epsilon R$ and \\
\begin{eqnarray}\label{eta_R}
R e^{i\psi} = \frac{1}{\langle k \rangle} \int \int \int k P(k)g(\omega)\eta(k,\omega,\theta,t) e^{i\theta} d\theta d\omega dk \nonumber \\
\end{eqnarray}
The flow velocity function $v(t)$ is defined by,\\
\begin{eqnarray}\label{velocity_func}
v(t)=\frac{d\theta}{dt}&=&\omega(t)+\lambda r k R' \sin(\psi-\theta-\alpha) \nonumber \\
&=& \omega(t)+\epsilon \lambda r k R \sin(\psi-\theta-\alpha)
\end{eqnarray}
Now by substituting (\ref{density_perturbation}),(\ref{eta_R}),(\ref{velocity_func}) into (\ref{continuity}) and neglecting the higher order term of $\epsilon$, we have\\
\begin{eqnarray}\label{partial_eta}
\frac{\partial \eta}{\partial t}=-\omega \frac{\partial \eta}{\partial \theta}+\frac{\lambda r k R \cos(\psi-\theta-\alpha)}{2\pi}
\end{eqnarray}
Now Fourier series expansion in complex form of the function $\eta(k,\theta,\omega,t)$ is given by,
\begin{eqnarray}\label{eta_fourier}
\eta(k,\theta,\omega,t)=c(k,\omega,t)e^{i\theta}+c^*(k,\omega,t)e^{-i\theta}+\eta^\bot(k,\theta,\omega,t)\nonumber \\
\end{eqnarray}
where $\eta^\bot$ represents the higher order Fourier harmonic terms. Now,\\
\begin{eqnarray}\label{rcos_1st}
Re^{i(\psi-\theta-\alpha)}=e^{-i(\theta+\alpha)}Re^{i \psi} &=& \frac{e^{-i(\theta+\alpha)}}{\langle k \rangle}\int \int \int k P(k)g(\omega)\eta(k,x,\omega,t) e^{i x} d\omega dx dk \nonumber \\
&=&\frac{2 \pi e^{-i(\theta+\alpha)}}{\langle k \rangle}\int \int k P(k)g(\omega) c^*(k,\omega,t) d\omega dk
\end{eqnarray}
Similarly we can write,\\
\begin{eqnarray}\label{rcos_2nd}
Re^{-i(\psi-\theta-\alpha)}=\frac{2 \pi e^{i(\theta+\alpha)}}{\langle k \rangle}\int \int k P(k)g(\omega) c(k,\omega,t) d\omega dk \nonumber \\
\end{eqnarray}
Now combining the Eq. \ref{rcos_1st} and Eq. \ref{rcos_2nd} we can get,\\
\begin{eqnarray}\label{rcos}
R\cos(\psi-\theta-\alpha) =\frac{\pi}{\langle k \rangle}\bigg[e^{-i(\theta+\alpha)}\int \int k P(k)g(\omega) c^*(k,\omega,t) d\omega dk +e^{i(\theta+\alpha)}\int \int k P(k)g(\omega) c(k,\omega,t) d\omega dk \bigg]
\end{eqnarray}
Using (\ref{eta_fourier}), (\ref{rcos}) into (\ref{partial_eta}) and compairing the coefficients of $e^{i\theta}$:
\begin{eqnarray}\label{c_eqn}
\frac{\partial c(k,\omega,t)}{\partial t}=-i \omega c(k,\omega,t) +\frac{\lambda r k}{2 \langle k \rangle}e^{i \alpha}\int_{k_{min}}^{\infty} \int_{-\infty}^{\infty} k' P(k')g(\omega') c(k',\omega',t) d\omega' dk'
\end{eqnarray}
Now for seperable of solution we can let $c(k,\omega,t)=A(k)B(\omega)e^{\mu t}$ and put it in Eq \ref{c_eqn}:\\
\begin{eqnarray}\label{putting_c}
\mu A(k)B(\omega)=-i \omega A(k)B(\omega) +\frac{\lambda r k}{2 \langle k \rangle}e^{i \alpha}\int_{k_{min}}^{\infty} \int_{-\infty}^{\infty} k' P(k')g(\omega')A(k')B(\omega') d\omega' dk'
\end{eqnarray}
Now if we let, \\
$\frac{\lambda r}{2 \langle k \rangle}e^{i \alpha}\int_{k_{min}}^{\infty} \int_{-\infty}^{\infty} k' P(k')g(\omega')A(k')B(\omega') d\omega' dk'=X$, then from (\ref{putting_c}), $A(k)B(\omega)=\frac{Xk}{\mu+i\omega}$, we have from above equations $A$ and $B$ are cancelled:\\
\begin{eqnarray}\label{forward_main}
1=\frac{\lambda r \langle k^2 \rangle}{2 \langle k \rangle}e^{i \alpha}\int \frac{(\mu-i\omega)g(\omega)}{\mu^2 + \omega^2}d\omega
\end{eqnarray}
When natural frequency $\omega$ follow a symmetric uniform distribution within an interval $[-a,a]$ then we have:
\begin{eqnarray}
g(\omega)&=& \frac{1}{2a}~~;~~ -a\leq\omega\leq a \nonumber \\
&=&0~~~~;~~ otherwise
\end{eqnarray}
From (\ref{forward_main}) we get,\\
\begin{eqnarray}
1=\frac{\lambda r \langle k^2 \rangle}{2a\langle k \rangle}e^{i\alpha} \tan ^{-1}(\frac{a}{\mu})
\end{eqnarray}
Now for marginal stability we can take $\mu \rightarrow 0$ in the above equation and we get,\\
\begin{eqnarray}
\lambda_f=\frac{4a\langle k \rangle \cos \alpha}{\pi r \langle k^2 \rangle}
\end{eqnarray}

\subsection{Impact of multiplexity on frustration:  validation in SF-SF multiplex networks}
We have shown in the main text that if a frustrated layer is unable to show ES , it may show ES in multilayer configuration (See Fig.\ (4) in main text where two ER networks are considered for $\alpha=0$ and $\alpha=1$). We have extended our result for SF-SF multilayer system in which one layer is frustrated where as other layer is not frustrated ($\alpha=0$). Fig.\ \ref{simu:sf_sf_N2000_alpha0_1} (b) shows that $(1)$ ES can be established for both layers if there is layer-layer interaction
and $(2)$ the order parameter   can be slightly enhanced with an emergence of  short window of explosive synchronization in multiplex network (shown in blue color). 

\end{widetext}

\begin{thebibliography}{28}
\bibitem{Boccaletti2002} S. Boccaletti, J. Kurths, G. Osipov, D.L. Valladares, C.S. Zhou, \textit{ Phys. Rep.} {\bf 366}, 1 (2002).
\bibitem{Pikovsky2003} A. Pikovsky, M. Rosenblum, and J. Kurths, \textit{Synchronization:
A Universal Concept in Nonlinear Sciences}, (Cambridge University Press, Cambridge, England, 2003). 
\bibitem{Strogatz2003} S. H. Strogatz, \textit{Sync: The Emerging Science of Spontaneous Order}, (Hypernion, New York 2003).


\bibitem{Kuramoto1984} Y. Kuramoto, \textit{Chemical Oscillations, Waves, and Turbulence}, (Springer, New York, 1984).
\bibitem{Bonilla2005} J. A. Acebr{\`o}n, L. L.  Bonilla, C. J. P. Vicente, F. Ritort, and R. Spigler, \textit{ Rev. Mod. Phys.} {\bf 77}, 137-185 (2005).

\bibitem{Pazo2005} D. Pa\'{z}o, \textit{Phys. Rev. E} {\bf 70}, 046211 (2005).

\bibitem{Kurths2016}F. A. Rodrigues, T. K. D. M. Peron, P. Ji, and J. Kurths, \textit{ Phys. Rep.} {\bf 610}, 1-98 (2016).

\bibitem{Boccaletti2016} S. Boccaletti, J.A. Almendral, S. Guan, I. Leyva, Z. Liu, I. Sendi{\~n}a-Nadal, Z. Wan, and Y. Zou, \textit{ Phys. Rep.} {\bf 660}, 1-94 (2016).

\bibitem{Jesus2011} J. G\'{o}mez-Garde\~{n}es, S. Gomez, A. Arenas, and Y. Moreno, \textit{Phys.Rev. Lett.} {\bf 106}, 128701 (2011).

\bibitem{Leyva2013} I. Leyva, A. Navas, I. Sendi\~{n}a-Nadal, J. A. Almendral,  J. M. Buldu, M. Zanin, D. Papo, and S. Boccaletti, \textit{Scientific Reports} {\bf 3} 1281 (2013).  Y. Zou, T. Pereira, M. Small, Z. Liu, and J. Kurths,  \textit {Phys. Rev. Lett.} {\bf 112} 114102 (2014).

\bibitem{Wang2017} C.Q. Wang, A. Pumir, N. B. Garnier, and  Z.  Liu, \textit{Front. Phys.} {\bf 12}(5), 128901 (2017).

\bibitem{Lee2018} U. Lee, M. Kim, K. Lee, C M. Kaplan, D. Clauw, S. Kim, G. Mashour, and  R. Harris, \textit{Scientific Reports} {\bf 8} 243 (2018).

\bibitem{Kumar2015}P Kumar, D  Verma, P. Parmananda, and S. Boccaletti, \textit{Phys. Rev. E} {\bf 91}, 062909 (2015).
\bibitem{Pinto2015} R. S. Pinto and A. Saa, \textit{Phys. Rev. E} {\bf 91}, 022818 (2015).

\bibitem{Zhang2013} X. Zhang, X. Hu, J. Kurths, and Z. Liu, \textit{Phys. Rev. E} {\bf 88}, 010802(R) (2013).

\bibitem{Leyvaweighted2013} I. Leyva, I. Sendi\~{n}a-Nadal,  J. A. Almendral, A. Navas, S. olmi, and S. Boccaletti, \textit{Phys. Rev. E} {\bf 88}, 042808 (2013).

\bibitem{Xu2016} C. Xu, Y. Sun, J. Gao, T. Qiu, Z. Zheng, and S. Guan, \textit{Scientific Reports} {\bf 6} 21926 (2016). 

\bibitem{Filatrella2007} G. Filatrella, N.F Pedersen, and K.Wiesenfeld, \textit{Phys. Rev. E} {\bf 75}, 017201 (2007).

\bibitem{Zhang2015} X Zhang, S Boccaletti, S Guan, and Z Liu,  \textit {Phys. Rev. Lett.} {\bf 114}(3), 038701 (2015).

\bibitem{Danziger2016} M M. Danziger, O. I. Moskalenko, S. A. Kurkin, X. Zhang, S. Havlin, and S. Boccaletti, \textit{Chaos} {\bf 26} 065307 (2016).

\bibitem{Kachhvah2017} A. Kachhvah and S. Jalan, \textit{Euro Phys. Lett.} {\bf 119} 60005 (2017).

\bibitem{Nicosia2017}V. Nicosia, P. S. Skardal, A. Arenas, and V. Latora, \textit{ Phys. Rev. Lett.} {\bf 119}, 138302 (2017).

\bibitem{Domenico2016} M. De Domenico, C. Granell, M A. Porter, and A. Arenas \textit{Nature Physics} {\bf 12}, 901–906 (2016).

\bibitem{Gomez2013} S. G\'{o}mez, A. D\'{i}az-Guilera, J. G\'{o}mez-Garde\~{n}es, C. J. P\'{e}rez-Vicente, Y. Moreno, and A. Arenas {\textit Phys. Rev. Lett.} {\bf 110}, 028701 (2013).

\bibitem{Sakaguchi1986} H. Sakaguchi and Y. Kuramoto, \textit{Prog. Theor. Phys.} {\bf 76}, 576 (1986).

\bibitem{Coutinho2013} B. C. Coutinho, A. V. Goltsev, S. N. Dorogovtsev, and J. F. F.
Mendes, \textit{Phys. Rev. E} {\bf 87}, 032106 (2013).
\bibitem{Ichinomiya2004} T. Ichinomiya, \textit{Phys. Rev. E} {\bf 70}, 026116 (2004).

\bibitem{Kundu2017} P. Kundu, P. Khanra, C. R. Hens, and P. Pal, \textit{Phys. Rev. E} {\bf 96}, 052216 (2017), P. Kundu,  C.  Hens, B. Barzel, and  P. Pal, \textit{Europhys. Lett.} {\bf 120}, 40002 (2017). 
\bibitem{Wu2018} H. Wu and M. Dhamala, \textit{arXiv}  1805.03510v1, (2018).
\bibitem{English2016} L. Q. English, D. Mertens, S. Abdoulkary, C. B. Fritz, K. Skowronski, and P. G. Kevrekidis, \textit{Phys. Rev. E} {\bf 94}, 062212 (2016). 
\bibitem{SI} See Supplementary material.

\end{thebibliography}

\begin{thebibliography}{28}
\bibitem{Ichinomiya2004} T. Ichinomiya, \textit{Phys. Rev. E} {\bf 70}, 026116 (2004).
\bibitem{Kundu2017} P. Kundu, P. Khanra, C. R. Hens, and P. Pal, \textit{Phys. Rev. E} {\bf 96}, 052216 (2017). P. Kundu,  C.  Hens, B. Barzel, and  P. Pal, \textit{Europhys. Lett.} {\bf 120}, 40002 (2017).
\end{thebibliography}
\end{document}